\documentclass[amsmath, amssymb, reprint, aps]{revtex4-1}
\usepackage{ulem}
\usepackage{bm}
\usepackage{makecell}
\usepackage{graphicx}
\usepackage{hyperref}
\hypersetup{
    colorlinks   = true, 
    urlcolor     = blue, 
    linkcolor    = black, 
    citecolor   = red 
}

\newcommand{\xv}{\textbf{x}}

\newcommand{\dv}{\textbf{d}}
\newcommand{\donev}{\textbf{d}_1}
\newcommand{\dtwov}{\textbf{d}_2}
\newcommand{\tv}{\bm{\theta}}
\newcommand{\tML}{\tv^{\rm ML}}

\begin{document}
    \title{A new measure of tension between experiments}    
    \author{Saroj Adhikari} \email{saroj@umich.edu}
    \author{Dragan Huterer} \email{huterer@umich.edu}    
    \affiliation{Department of Physics, University of Michigan, 450 Church St, Ann Arbor, MI 48109-1040;\\ 
        Leinweber Center for Theoretical Physics, University of Michigan, 
        450 Church St, Ann Arbor, MI 48109-1040}
    \date{\today}
    
    \begin{abstract}
        Tensions between cosmological measurements by different surveys or probes have always been important --- and are presently much discussed --- as they may lead to evidence of new physics.  Several tests have been devised to probe the consistency of datasets given a cosmological model, but they often have undesired features such as dependence on the prior volume, or burdensome requirements such as that of near-Gaussian posterior distributions. We propose a new quantity, defined in a similar way as the Bayesian evidence ratio, in which these undesired properties are absent. We test the quantity on simple models with Gaussian and non-Gaussian likelihoods. We then apply it to data from the {\it Planck} satellite: we investigate the consistency of $\Lambda$CDM model parameters obtained from TT and EE angular power spectrum measurements, as well as the mutual consistency of cosmological parameters obtained from large scale (multipoles, $\ell<1000$) and small scale ($\ell \geq 1000$) portions of each measurement and find no significant discrepancy in the six-dimensional $\Lambda$CDM parameter space.  
    \end{abstract}
    
    \maketitle
    \medskip
    \textit{Introduction.}
    The use of Bayesian statistics in cosmology is now commonplace: most of the results on cosmological parameters from cosmic microwave background (CMB) experiments \cite{Ade:2015xua} and large-scale structure (LSS) surveys \cite{Abbott:2017wau} are reported as posterior distributions. In addition, various Bayesian methods are used for model comparison \cite{Trotta:2008qt}.
    
    Along with the increase in the number of cosmological surveys and the improvement in their precision, a number of tensions between  parameters derived from different experiments have been observed. For example, the Hubble constant $H_0$ measured using the distance ladder in the local universe disagrees with that derived from {\it Planck} CMB observations; in the standard six parameter $\Lambda$CDM model, the disagreement is about $3.4\sigma$ \cite{Riess:2016jrr, Bernal:2016gxb}. There is also some tension between the measurements of the amplitude of fluctuations $\sigma_8$ and the matter density $\Omega_m$ from weak lensing to that of the measurement from {\it Planck} CMB data \citep{Hildebrandt:2016iqg,Troxel:2017xyo,Troxel:2018qll}. As a result, a number of statistics have been developed to compare datasets in cosmology. The primary goal of these statistics is to determine if two datasets are consistent realizations of the same model, that is, with a single set of cosmological parameters (see \cite{Seehars:2015qza, Charnock:2017vcd, Lin:2017ikq} for discussions and comparisons of some of the popular methods). For an alternative approach using hyperparameters, see \cite{Hobson:2002zf, Bernal:2018cxc}.
    
    The Bayesian evidence-based metric of \cite{Marshall:2004zd} has been widely used \cite{March:2011rv, Amendola:2012wc, Martin:2014lra, Joudaki:2016mvz, Raveri:2015maa}, but is known to strongly depend on the priors given to parameters. This has led to the use of other measures \cite{Seehars:2014ora,Grandis:2016fwl, Feeney:2018mkj} that do not have the prior-volume dependence, but at the expense of losing the simplicity of an evidence ratio. 
    In this work, we define an evidence-based quantity which fixes the problem of prior volume dependence and which can be evaluated on a easy-to-interpret scale.
    
    Consider two datasets $\dv_1$ and $\dv_2$, and let $\tv$ be the  parameters of a model. Let us assume that both the datasets and the combination of them can be modeled by a particular $\Lambda$CDM realization, and the priors are wide enough to include the parameter posteriors preferred by both the datasets individually. Most commonly, a Bayesian analysis is used to determine posterior probability distributions for the model parameters $\tv$. Suppose the two datasets separately give two (normalized) posterior distributions
    \begin{equation}
    \!\!\!
    p_1(\tv | \dv_1) = \frac{ \mathcal{L}(\dv_1|\tv) \pi(\tv)}{E(\dv_1)}; \,\,\,
    p_2(\tv | \dv_2) = \frac{\mathcal{L}(\dv_2|\tv) \pi(\tv)}{E(\dv_2)}, 
    \label{eq:p1p2}
    \end{equation}
    where $\mathcal{L}(\dv|\tv)$ denotes the likelihood of the data $\dv$ given the model defined by a set of parameters $\tv$, $\pi(\tv)$ is the prior probability of the model parameters, and $E$ is called the marginal likelihood or the evidence,
    \begin{align}
        E(\dv) &= \int d\tv \mathcal{L}(\dv|\tv) \pi(\tv).
        \label{eq:evidence}
    \end{align}
    
    We will always use normalized probability density functions for the likelihood ($\int \mathcal{L}(\dv|\theta) d\dv = 1$) and the prior  ($\int \pi(\tv) d \tv =1$). The posterior for the combination of datasets $\dv_1, \dv_2$ is:
    \begin{equation}
    \begin{aligned}
    \!\!\!
    p_{12}(\tv|\dv_1,\dv_2) = \frac{\mathcal{L}(\dv_1, \dv_2, \tv) \pi(\tv)}{E(\dv_1, \dv_2)}   = \frac{\mathcal{L}(\dv_1, \tv) \mathcal{L}(\dv_2, \tv) \pi(\tv)}{E(\dv_1, \dv_2)},
    \end{aligned}
    \end{equation}
    where the second equality assumes that the combined likelihood is approximated well by the product of the two likelihoods.
    
    The ratio of the evidences obtained using two different models, called the Bayes Factor, is a widely used measure for model comparison. In this work, we will define a similar ratio to compare two sets of parameter constraints of a model obtained using different datasets or experiments. 
    
\begin{figure}
        \includegraphics[width=0.48\textwidth]{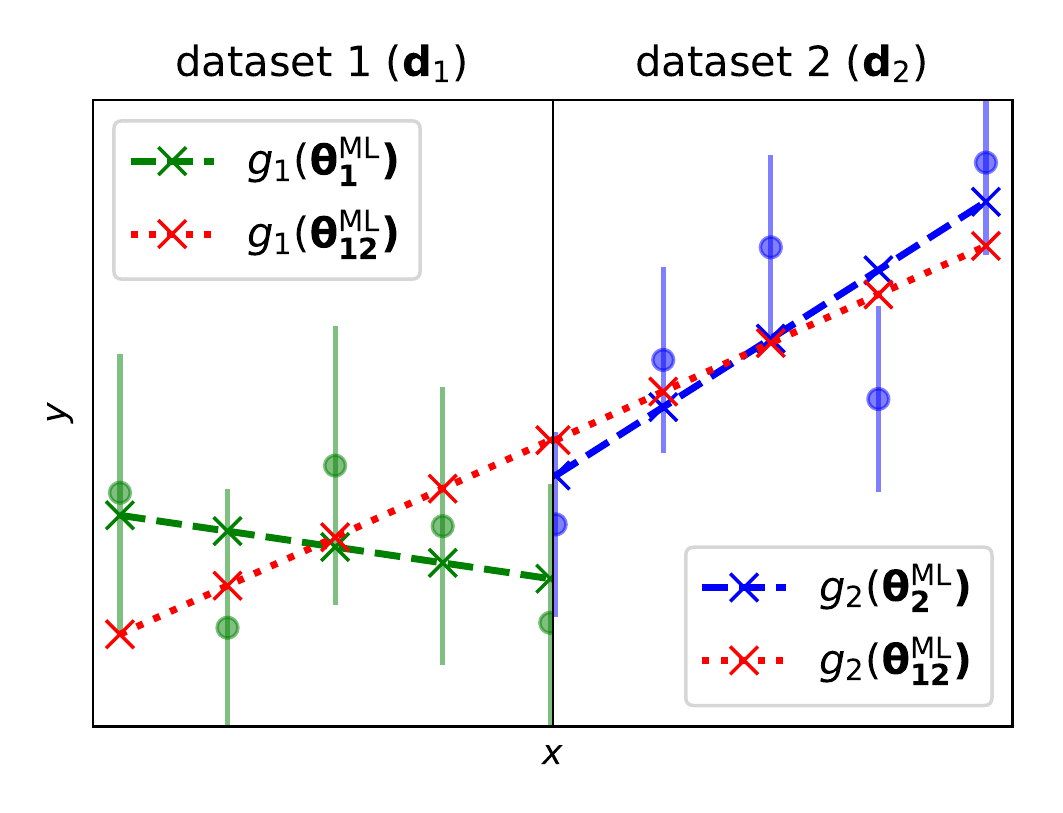}
        \caption{Illustration of how we make use of the maximum-likelihood data-realizations (crosses) as opposed to the actual measurements (dots) in our evidence integrals. Doing this gets rid of the data scatter and therefore makes our statistic only sensitive to the model parameters and not the spread in the data realizations.}
        \label{fig:gtheta}
    \end{figure}

    \medskip
    \textit{Evidence for model parameters.}
    We first define the marginal likelihood (or the evidence) for the maximum likelihood model parameters, $\tML$, instead of the usual definition of the evidence for the data, $\dv$. We do so as our primary goal is to quantify the level of consistency between model parameters obtained from different datasets or experiments. Analogous to Eq.~(\ref{eq:evidence}), we define the evidence for the maximum likelihood model parameters
    \begin{align}
        E(g(\tML)) &= \int d\tv \mathcal{L}(g(\tML)|\tv) \pi(\tv),
    \end{align}
    where, instead of the measured data, we have used the maximum likelihood values of the data realization given the model, $g(\tML)$; see Figure \ref{fig:gtheta} for an illustration. Here $g(\tv)$ is the function that computes the model prediction for the data given the parameters $\tv$; for example, in the case of the CMB temperature fluctuation data, the model prediction is represented by the theory angular power spectra $g_{\rm cmb}^{TT}(\tv) = \{C_\ell^{TT}\}$. If the likelihood in the above equation is a combination of two experiments, then we can define evidences for the maximum likelihood parameters obtained through the combination of the two datasets (denoted by $i,j$), $E(g(\tML_{ij}))$. Alternatively, we can define an evidence so that each part of the data vector in the evidence integral uses its own maximum likelihood parameter values, obtained by analyzing each experiment separately, $E(\{g_i(\tML_i), g_j(\tML_j)\})$. As we will show, the ratio of the two evidences can quantify the tension between the parameter constraints obtained from two different datasets.
    
    \medskip
    \textit{Evidence-based dataset comparison.}
    For simplicity, consider two datasets that have independent likelihoods, $\mathcal{L}_i(\dv|\tv), \mathcal{L}_j(\dv|\tv)$, and let the measured data vector for each experiment be denoted by $\dv_i, \dv_j$. Let us further assume that the maximum likelihood parameters of the model, $\Lambda$CDM for example, are known for three different cases: two datasets analyzed separately, $\tML_i, \tML_j$, and their combined analysis, $\tML_{ij}$.
    
    Our null hypothesis $\mathcal{H}_0$ is that both the datasets are realizations of a single set of parameters, $\tML_{ij}$, from the combined fit. The alternative, more complicated, hypothesis $\mathcal{H}_1$ is that each of the datasets are realizations of their own set of parameters, $\tML_i, \tML_j$. Then, using the Bayes theorem similarly to the derivation of the Bayes factor, we get
    \begin{equation}
    \begin{aligned}
    \frac{p(\mathcal{H}_1)}{p(\mathcal{H}_0)} &= \frac{\int d\tv \pi(\tv) \mathcal{L}_i(g_i(\tML_i)|\tv) \mathcal{L}_j(g_j(\tML_j)|\tv)}{\int d\tv \pi(\tv) \mathcal{L}_i(g_i(\tML_{ij})|\tv) \mathcal{L}_j(g_j(\tML_{ij})|\tv)} \label{eq:DER} \\
    &= \frac{E_{ij}^{\rm sep}}{{E_{ij}^{\rm com}}} \equiv R_{ij},
    \end{aligned}
    \end{equation}
    where the superscripts \texttt{sep} and \texttt{com} in the formula above stand for separate and combined maximum likelihood parameters, respectively. We will first consider a case where the above expression can be evaluated analytically.
    
    Consider two likelihoods given by two $N_{\rm params}$-dimensional multivariate Gaussian distributions with arbitrary covariance matrices $\Sigma_1$ and $\Sigma_2$,
    \begin{align}
        \mathcal{L}_1(\tv) =& \frac{1}{\sqrt{\det(2\pi \Sigma_1)}} \exp\left[{-\frac{1}{2}}(\donev-\tv)^T \Sigma_1^{-1} (\donev-\tv)\right] \nonumber \\
        \mathcal{L}_2(\tv) =& \frac{1}{\sqrt{\det(2\pi \Sigma_2)}} \exp\left[{-\frac{1}{2}}(\dtwov-\tv)^T \Sigma_2^{-1} (\dtwov-\tv)\right].\nonumber
    \end{align}
    
    In this simple example, we have taken $g_{i,j}(\tv) = \tv$ so that the expressions are easy to evaluate analytically. If we further assume that the prior on each of the parameters is uniform and wide (compared to the constraint on the parameter), we get \cite{IMM2012-03274},
    \begin{align}
        \nonumber
        E_{12}^{\rm sep} &\propto \int d\tv \mathcal{L}_1(\dv_1|\tv) \mathcal{L}_2(\dv_2|\tv) 
        = \frac{1}{\sqrt{\det\left[2\pi(\Sigma_1+\Sigma_2)\right]}} \\ & \times\exp \left[-\frac{1}{2}(\donev-\dtwov)^T(\Sigma_1+\Sigma_2)^{-1}(\donev-\dtwov)\right] 
        \\[0.1cm]
        E_{12}^{\rm com} &\propto \int d\tv \mathcal{L}_1(\dv_{12}|\tv) \mathcal{L}_2(\dv_{12}|\tv)  
        = \frac{1}{\sqrt{\det\left[2\pi(\Sigma_1 + \Sigma_2)\right]}}
        \nonumber
    \end{align}
    so that
    \begin{align}
        \!\!\!
        R_{12} =& \exp \left[-\frac{1}{2}(\donev-\dtwov)^T(\Sigma_1+\Sigma_2)^{-1}(\donev-\dtwov)\right], \label{eq:R12NGaus}
    \end{align}
    the {negative} logarithm of which ($-\ln R_{12}$) is the two-experiment index of inconsistency (IOI) defined in \cite{Lin:2017ikq}. Under these conditions assuming that the null hypothesis is true, $(-2 \ln R_{ij})$ is $\chi^2$ distributed with $N_{\rm params}$ degrees of freedom (dof) \cite{Raveri:2018wln} (see their definition and discussion of $Q_{\rm DM}$). More generally, the ratio of probabilities of two hypotheses (evidence ratio) is similar to a likelihood-ratio test \cite{kassr95}, and the distribution of $(-2 \ln R_{ij})$ asymptotically approaches $\chi^2_{N_{\rm params}}$ by Wilks theorem \cite{Wilks}. Here, $N_{\rm params}={\rm dof}(\mathcal{H}_0)-{\rm dof}(\mathcal{H}_1)$, when comparing two datasets. We will, therefore, evaluate the probability-to-exceed (PTE) value of observed $\ln R_{ij}$ values by taking $(-2 \ln R_{ij})$ to be $\chi^2_{N_{\rm params}}$ distributed. 
    
    For two one-dimensional Gaussian likelihoods: $\mathcal{L}_1 = \mathcal{N}(d_1, \sigma_1)$ and $\mathcal{L}_2 = \mathcal{N}(d_2, \sigma_2)$, we get $(-2\ln R_{12})=(d_1-d_2)^2/(\sigma_1^2+\sigma_2^2)$. The application of our new measure to the marginalized Hubble constant $H_0$ likelihoods from {\it Planck} $\mathcal{L}_1 \sim \mathcal{N}(66.93, 0.62)$ \cite{Ade:2015xua} and distance ladder $\mathcal{L}_2 \sim \mathcal{N}(73.24, 1.74)$ \cite{Riess:2016jrr}, therefore, trivially gives us the values expected from Gaussian statistics i.e $\ln R_{12} = -5.83$ \cite{Lin:2017bhs} with a p-value $6.4\times 10^{-4}$ or $3.4\sigma$.
    
\begin{figure}
        \centering
        \includegraphics[width=0.48\textwidth]{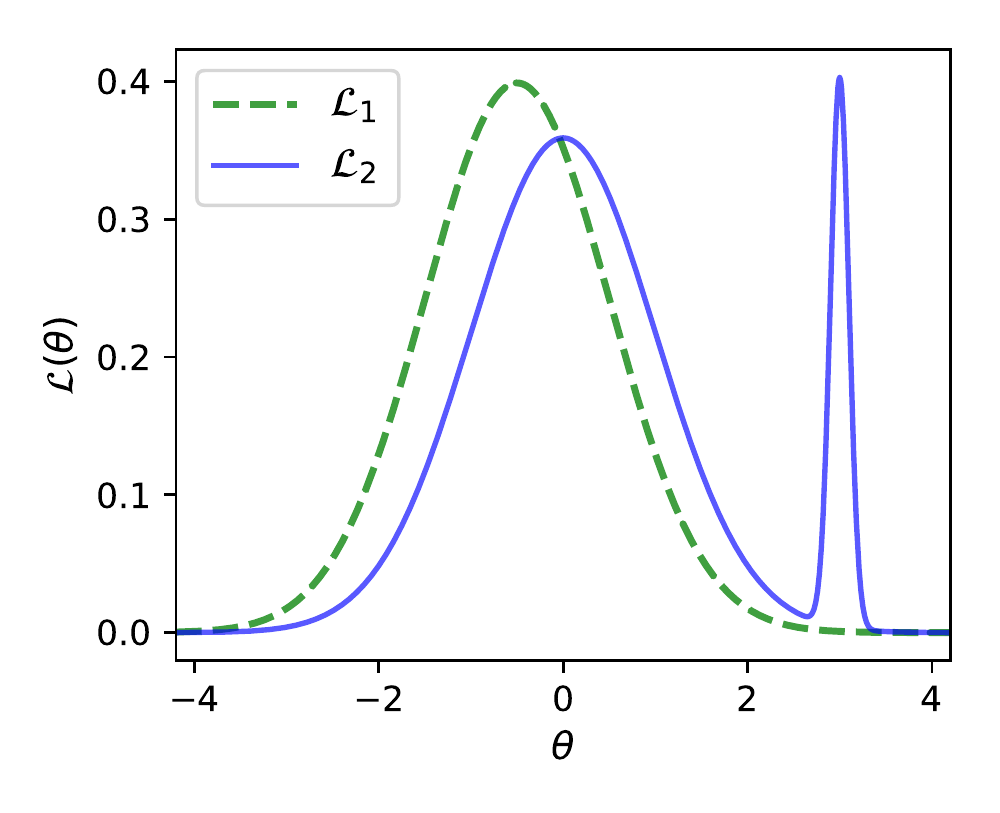}
        \caption{In this example, we verify that the two distributions shown in the figure: (i) Gaussian (dashed), and (ii) non-Gaussian (solid), are consistent with each other; the effect of the peak around $\theta=3$ in the non-Gaussian distribution is small with $\ln R_{12}=-0.064$ (compared to $\ln R_{12}=-0.0625$ if $\mathcal{L}_2$ had no second peak at $\theta=3$).}
        \label{fig:example1}
    \end{figure}

    Also, we note that our new measure is related to the \textit{tension} measure defined in \cite{Verde:2013wza}, because in some situations $E_{ij}^{\rm com}$ can be approximated by shifting one of the posterior probability density functions while preserving its shape. However, there can be ambiguity in the process of shifting one or both of the posterior distributions (for non-Gaussian and multimodal distributions), as discussed in Section X.B. of \cite{Lin:2017ikq}. That ambiguity is removed in our definition, as we reference the likelihood functions directly. We provide an example in Figure \ref{fig:example1}, in which the Gaussian likelihood is simply, $\mathcal{L}_1(d|\theta) = \mathcal{N}(d, 1)$. The non-Gaussian likelihood is a (normalized) sum of two Gaussians, defined as $\mathcal{L}_2(d|\theta) = 0.9 \mathcal{N}(d, 1) + 0.1 \mathcal{N}(d+3, 0.1)$. The distributions plotted in Figure \ref{fig:example1} are $\mathcal{L}_1(d=-0.5|\theta)$ (dashed) and $\mathcal{L}_2(d=0|\theta)$ (solid). Because the combined fit is insensitive to the narrow peak near $\theta=3$, we get $\ln R_{12}=-0.064$ without any ambiguity in how to shift the distributions, which shows that the two sets of parameters $\theta_1=-0.5$ and $\theta_2=0$ from the two likelihoods are consistent, as expected. Without the additional peak at $\theta=3$, the level of consistency is slightly better: $\ln R_{12} = -0.5\times(0.5^2)/2 = -0.0625$, a simple verification that the new measure gets contribution from non-Gaussian features.
    
    Next, we calculate $\ln R_{12}$ using different pairs of datasets (e.g. TT vs EE) from the {\it Planck} satellite, in which case $g_i(\tv)$ is no more a simple linear function but has to evaluated numerically.
    
    \medskip
    \textit{Application to Planck data.} 
    We use the binned and foreground-marginalized \texttt{plik\_lite} likelihood from the Planck collaboration \cite{Aghanim:2015xee} which includes multipoles $30-2508$ for TT power spectrum, and multipoles $30-1996$ for EE power spectrum. We fix the {\it Planck} calibration factor $y_p$ to 1; see Sec.\ C.6.2 of \cite{Aghanim:2015xee}, from which the CMB-only Gaussian \texttt{plik\_lite} likelihood is:
    \begin{align}
        \ln \mathcal{L}(\tilde{C}_b^{\rm CMB}|C_b^{\rm th}) = -\frac{1}{2}\xv^T \tilde{\Sigma}^{-1} \xv -\frac{1}{2} \ln \left[ \det(2\pi \tilde{\Sigma})\right], \label{eq:PlancklnL}
    \end{align}
    where $\xv = \tilde{C}_b^{\rm CMB}/y_p^2 - C_b^{\rm th}$. The binned and marginalized mean $\tilde{C}_b^{\rm CMB}$ and covariance matrix $\tilde{\Sigma}$ are provided by the {\it Planck} team. To evaluate the likelihood in Eq.~(\ref{eq:PlancklnL}), we compute lensed $C_\ell^{\rm th}$ for a given set of parameters $\tv$ using \texttt{camb} \cite{Lewis:1999bs,Howlett:2012mh} and bin the $C_\ell^{\rm th}$ using the appropriate weights to get $C_b^{\rm th}$.
    
    Without low-multipole polarization data, the optical depth to reionization $\tau$ is only weakly constrained and is strongly degenerate with the amplitude of scalar fluctuations $A_s$. To break this degeneracy, we use a low-$\ell$ polarization prior $\tau = 0.07\pm0.02$. The evidences we compute are:
    \begin{equation}
    \begin{aligned}
    \!\!\!
    E^{\rm sep}_{\rm TT, EE} =& \int d\tv\, \pi(\tv)  
    \mathcal{L}(\{C_\ell^{\rm TT}(\tML_T), C_\ell^{\rm EE}(\tML_E)\}|\tv)  \\
    \!\!\!
    {E}^{\rm com}_{\rm TT, EE} =& \int d\tv\, \pi(\tv)  
    \mathcal{L}(\{C_\ell^{\rm TT}(\tML_C), C_\ell^{\rm EE}(\tML_C)\}|\tv)
    \end{aligned}
    \end{equation}
    where $\tML_T$ and $\tML_E$ are obtained individually by using the respective TT and EE data, while $\tML_C$ is the maximum likelihood model parameters from the combined fit. We obtain the maximum likelihood values $\tML$ by using a global optimization algorithm {\tt differential\_evolution} \cite{Storn1997} implemented in {\tt scipy} \cite{scipy}. We calculate the evidences using the \texttt{MultiNest} package \cite{Feroz:2007kg, Feroz:2008xx}, and quote results and statistical errorbars produced by the importance nested sampling method \cite{Feroz:2013hea}. For evidence calculations, we take uniform priors on six cosmological parameters listed in Table \ref{table:priors}.
    \begin{table}
        \centering
        \caption{Cosmological parameters and their prior ranges.  $A_s$ and $n_s$ are the amplitude and spectral index of primordial scalar fluctuations, $\Omega_c$ and $\Omega_b$ are cold dark matter and baryonic matter densities, $H_0$ is the Hubble constant, and $\tau$ is the optical depth to reionization.} \label{table:priors}
        \begin{tabular*}{0.48\textwidth}{c@{\extracolsep{\fill}} r  c r}
            \hline \hline
            Parameter & Range & Parameter & Range \\ \hline
            $\ln(10^{10}A_s)$ & [2.7, 3.4] & $n_s$ & [0.8, 1.2] \\
            $\Omega_c$ & [0.1, 0.45] & $\Omega_b$ & [0.044, 0.055] \\
            $H_0$ & [50, 95] & $\tau$ & [0.005, 0.2] \\
            \hline \hline
        \end{tabular*}
    \end{table}
    
    The results are shown in Table \ref{table:lnRTEC} where, in addition to $\ln R$, we also quote the corresponding probability-to-exceed (p-value) and Gaussian $n$-$\sigma$ values. For the discrepancy between model parameters obtained from TT and EE spectra, we obtain $\ln R_{\rm TT,EE}=-1.93\pm 0.03$ (approximately $0.4 \sigma$). Previous studies also find no indication of strong discrepancy between these datasets \cite{Shafieloo:2016zga}, albeit by using more complicated methods, or by directly using the posteriors \cite{Lin:2017bhs}.
    
    \begin{table}[t]
        \centering
        \caption{Summary of applications of our new statistic to {\it Planck} data, discussed in the text. The second column shows $\ln R$ for TT and EE as the two datasets using {\it Planck} \texttt{plik\_lite} likelihood. The last two columns show $\ln R$ for splitting the data into two multipole ranges at $\ell_{\rm split}=1000$ using only TT or only EE data. Each p-value is computed by taking $(-2\ln R)$ as $\chi^2_{N_{\rm params}}$ distributed, and the corresponding Gaussian $n$-$\sigma$ value is also quoted in parentheses.}
        \label{table:lnRTEC}
        \begin{tabular*}{0.48\textwidth}{c@{\extracolsep{\fill}} c c  c}
            \hline \hline
            datasets & TT,EE & \thead{TTlow, TThigh\\$(\ell_{\rm split}=1000)$} & \thead{EElow, EEhigh\\$(\ell_{\rm split}=1000)$}\\
            \hline
            $\ln R$ & $-1.93 \pm 0.03$ & $-4.13 \pm 0.16 $ & $-0.83 \pm 0.16$ \\
            p-value & $0.7 (0.4\sigma) $ & $0.22 (1.2\sigma)$ & $0.95 (0.1\sigma)$\\
            \hline\hline
        \end{tabular*}
    \end{table}
    
    We perform another test using the {\it Planck} power spectrum data, by splitting the temperature data into $\ell<1000$ and $\ell \geq 1000$ samples and calculating $\ln R$ for these two datasets. We again find that the level of inconsistency is small with $\ln R_{\rm TTsplit} = -4.13 \pm 0.16$ or approximately $1.2 \sigma$, which agrees with the significance obtained using simulated data sets in \cite{Aghanim:2016sns}. Note that, to obtain the values in Table \ref{table:lnRTEC}, we are using the \texttt{plik\_lite} likelihood in which low-$\ell$ ($\ell<30$) multipoles are not included; inclusion of these large-scale multipoles would likely increase the discrepancy as their amplitude is known to be anomalously low.
    
    To estimate the effect of low-$\ell$ part of the TT likelihood, we implement an approximation to the low-$\ell$ likelihood following \cite{Aghanim:2016sns} (see their Section 3.2 for details), which they have tested to find that the approximation gives similar cosmological parameters compared to the computationally more demanding pixel-space likelihood. To summarize: $f_\ell(2\ell+1) \hat{C}_\ell/C_\ell$ is drawn from a $\chi^2_{f_\ell(2\ell+1)}$ probability distribution function, where $f_\ell$ are mask-dependent fitting factors determined for the {\tt commander} mask. Here $\hat{C}_\ell$ is the mask-deconvolved power spectrum, which we take to be the {\it Planck} {\tt commander} quadratic maximum likelihood (QML) $C_\ell$s. Any correlation between different multipoles for $\ell<30$ and with the {\tt plik\_lite} multipole bins is ignored. For $\ell_{\rm split}=1000$, including the approximate low-$\ell$ likelihood, we now get $\ln R_{\rm TTsplit}=-5.32 \pm 0.05$ or approximately $1.6\sigma$, which again agrees with the significance quoted in \cite{Aghanim:2016sns} obtained using simulations. 
    
    {We finally carry out a similar analysis with the polarization data: we split the {\it Planck} EE data in multipole, using the {\tt plik\_lite} likelihood for each multipole range}. The large and small scale multipole split for the EE spectrum results in consistent $\Lambda$CDM parameters: $\ln R_{\rm EEsplit}=-0.83\pm 0.16$, or approximately $0.1\sigma$, which is expected given the lesser constraining power of the EE spectrum for {\it Planck} noise levels.
    
    \medskip
    \textit{Summary and Conclusion.} \label{sec:summary}
    We have introduced a new statistic to quantify tension between experiments. The statistic is based upon Bayesian evidence, and has advantages of not depending on the prior volumes of the parameters, and of being straightforward to apply to multiparameter, non-Gaussian likelihood distributions. We have shown that our new measure reduces to the expected discrepancy measure for Gaussian distributed posteriors, and gives sensible results in the non-Gaussian tests that we performed. 
    
    Applying {the new statistic} to the {\it Planck} power spectrum data, we find  that the cosmological parameters obtained from TT and EE spectra are consistent, and that the level of discrepancy of the parameters obtained from the TT spectrum split into smaller and larger scales at $\ell_{\rm split}=1000$ is slightly larger at about $1.6 \sigma$.
    
    We have limited our application to just the {\it Planck} data in this work. It is worthwhile to apply the new measure to comparing the {\it Planck} constraints with weak-lensing constraints \cite{Abbott:2017wau} and smaller-scale CMB constraints \cite{Aylor:2017haa}. It will also be useful to consider using the statistic in the context of $\Lambda$CDM extensions. Further, we have only carefully investigated the ratio for comparing two datasets. A straightforward application of the ratio for more than two datasets might be possible by evaluating $(-2\ln R)$ as $\chi^2$ distributed with $N_{\rm params}\times(N_{\rm sets}-1)$ degrees of freedom, but detailed investigation of this possibility and application to other cosmological datasets is left for future study.
    
    \medskip
    \textit{Acknowledgments.}
    The authors are supported by NASA under contract 14-ATP14-0005. DH is also supported by DOE under
    Contract No. DE-FG02-95ER40899. This work used the Extreme Science and Engineering Discovery Environment (XSEDE), which is supported by National Science Foundation grant number ACI-1548562. We thank Marius Millea for providing the necessary $f_\ell$ coefficients and example code to implement the low-$\ell$ approximated likelihood. We are grateful to Wayne Hu, Marco Raveri, Vivian Miranda, Weikang Lin, Mustapha Ishak-Boushaki, Pavel Motloch and Michael Hobson for insightful comments.

\end{document}